\documentclass[a4paper,11pt]{article}
\usepackage{epsfig}

\title{Highlight: Forward Physics (LHCf + FASER)
\thanks{Presented at The Ninth Annual Conference on Large Hadron Collider Physics - LHCP2021, 7-12 June 2021, Online}
}

\author{E. Berti\footnote{eugenio.berti@fi.infn.it}
\\
\small University of Firenze, Via G. Sansone 1, Sesto Fiorentino, Firenze, Italy\\
\small INFN Section of Florence, Via B. Rossi 2, Sesto Fiorentino, Firenze, Italy\\
on behalf of the \textit{LHCf} and \textit{FASER} collaboration
}

\begin{document} 
\maketitle

\abstract{
The LHC Run III will be a crucial run for the two LHC forward experiments: LHCf and FASER. In particular, Run III will be the last run where the LHCf detector can operate and, at the same time, the first run of the new FASER project. The LHCf experiment is dedicated to precise measurements of forward production, necessary to tune hadronic interaction models employed in cosmic-ray physics. In Run III, the experiment will accomplish two fundamental goals: operating in p-p collisions at $\sqrt{s} = $ 14 TeV, it will acquire a ten times larger statistics respect to Run II, in order to have precise measurements of $\pi^{0}$ production; operating in high energy p-O and O-O collisions, it will measure forward production in a configuration that is very similar to the first interaction of an Ultra High Energy Cosmic Ray with an atmospheric nucleus. The FASER experiment is dedicated to the search of new weakly-interacting light particles, thanks to a forward detector with proper shielding from Standard Model background. In Run III, the experiment will search for new particles with a good sensitivity, which can be strongly improved by a following possible upgrade before HL-LHC. In addition, thanks to the dedicated FASER$\nu$ detector, the experiment will measure neutrino production at a collider for the first time. In this contribution, we discuss the main results expected from the LHCf and FASER experiments in Run III, highlighting their fundamental contribution in research fields that are not accessible to the four large LHC experiments.
}

\section{Introduction}

The term \textit{forward physics} indicates a very broad and variegated research field that includes experiments with quite different purposes. In the LHC Run III, two forward experiments will operate: the LHCf experiment, dedicated to precise forward production measurements that can be used to tune hadronic interaction models, and the FASER experiment, dedicated to the search of new weakly-interacting light particles and the measurements of neutrino production at TeV energies. In the following, we separately discuss the highlights expected by these two experiments in Run III.

\section{The LHCf experiment in LHC Run III}

The measurements of the average composition of Ultra High Energy Cosmic Rays (UHECR) is strongly affected by the large theoretical uncertainty associated to the hadronic interaction models that are used to simulate the interaction of an UHECR with the atmosphere. In order to reduce this uncertainty it is necessary to provide high energy calibration data that can be used for model tuning. The Large Hadron Collider is the most suitable place where we can perform these measurements, since a p-p collision at $\sqrt{s} = $ 14 TeV allows us to study a configuration that is equivalent to the first interaction of $10^{17}$ eV proton cosmic ray with an atmospheric proton at rest. In this configuration, the LHCf experiment \cite{ref:lhcf} gives fundamental information relative to forward production, by detecting secondary neutral particles (mainly neutrons, photons and, indirectly, $\pi^{0}$s) with pseudorapidity $\eta > 8.4$. This is possible thanks to two imaging and sampling calorimeter detectors \cite{ref:performance} that are installed in regions called TArget Neutral absorber (TAN), located at a distance of 141.05~m from Interaction Point 1 (IP1). In LHC Run I and II, LHCf acquired data relative to high energy p-p and p-Pb collisions, publishing several results on forward photon \cite{ref:photon7tev, ref:photon13tev}, $\pi^{0}$ \cite{ref:pi7tev, ref:piPb, ref:piAll} and neutron \cite{ref:neutron7tev, ref:neutron13tev, ref:neutron13tev_bisse} production. Due to the following reshaping of the TAN region, the detectors cannot operate in HL-LHC, but in Run III the LHCf experiment will reach its main scientific goals. This will be achieved by operating in two different configurations: proton-proton collisions at $\sqrt{s} = $ 14 TeV, and proton-oxygen and oxygen-oxygen collisions at $\sqrt{s_{NN}} = $ 5.52 or 7 $\times$ Z TeV.

\begin{figure}[tbp]
\centering
\begin{center}
\includegraphics[width=0.95\linewidth]{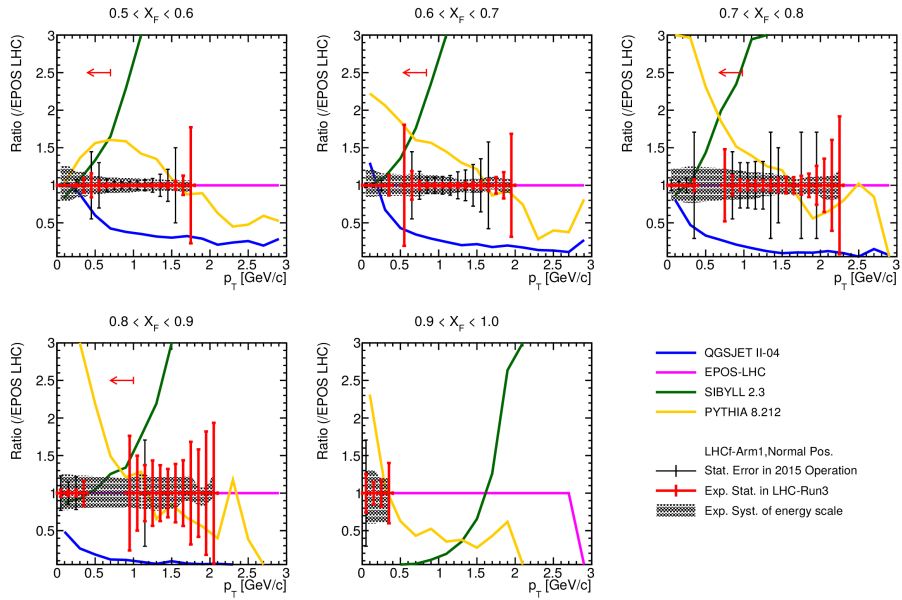}%
\end{center}
\caption{Expected reduction of statistical uncertainty on the measurements of forward $\pi^{0}$ production between the operations in Run II (black markers) and III (red markers). Gray area is the expected systematic uncertainty. It is clear the this reduction will give more precise information for model testing and tuning.
}
\label{fig:lhcf}
\end{figure}

The main motivation for p-p collisions at $\sqrt{s} = $ 14 TeV is to precisely measure forward production at the highest available energy. This is possible thanks to an upgrade of the Arm2 silicon readout electronics that increases the data acquisition rate by a factor ten. Thus, considering a couple of days of data taking, we expect to collect a statistics corresponding to an integrated luminosity of about 20~$\mathrm{nb^{-1}}$, ten times larger than the one acquired in p-p collisions at $\sqrt{s} = $ 13 TeV during Run II. As shown in Fig.\ref{fig:lhcf}, this large data sample is particularly important to strongly decrease the statistical uncertainty in the measurement of $\pi^{0}$ forward production, indirectly reconstructed by the simultaneous detection of the two $\gamma$s generated by the decay. 

The main motivation for high energy p-O and O-O collisions is due to the fact that these two configurations correspond to a situation very similar to the first interaction of an UHECR with an atmospheric nucleus, which is generally a light nucleus, like N or O. In Run I and II, the LHCf experiment measured forward production only in high energy p-p and p-Pb collisions, and the information on the realistic case where the proton interacts with a light nucleus must be obtained by interpolating between these two measurements. In Run III, thanks to p-O collisions, we will directly measure the change in forward production in a case of a light nucleus target, eliminating this interpolation uncertainty. O-O collisions are also very important for cosmic-ray application, since recent measurements of average composition qualitatively indicate that UHECRs are made of an admixture of light nuclei, but not protons only. More details can be found in \cite{ref:icrc}.

In LHC Run III, the LHCf experiment will have common operations with the ATLAS experiment \cite{ref:atlas}. The LHCf-ATLAS joint analysis proved to be a very powerful tool to separate different mechanisms responsible for forward production (diffractive and non-diffractive processes) by using a tag in the central region \cite{ref:lhcf-atlas}. In addition, two important ATLAS subdetectors will be included in common operations: the ALFA roman pots, in order to get more information for the identification of single-diffractive events and to study exclusive processes like the $\Delta$ resonance, and the ZDC, in order to significantly improve the energy resolution of hadronic shower from about 40 to 20\%.

\section{The FASER experiment in LHC Run III}

The FASER experiment \cite{ref:faser} is installed in the TI12 tunnel, originally built to connect SPS to LEP, at a distance of 480 m from IP1, after a rock and concrete shielding of approximately 100 m. As for the LHCf case, the FASER detector is a very simple and cheap detector with a strong physics potential. The detector consists in a decay volume of 1.5~m length, followed by a magnetic spectrometer and an electromagnetic calorimeter, with scintillator layers at different depths working as veto, trigger, timing or preshower stations \cite{ref:fasertp}. 

\begin{figure}[tbp]
\centering
\begin{center}
\includegraphics[width=0.90\linewidth]{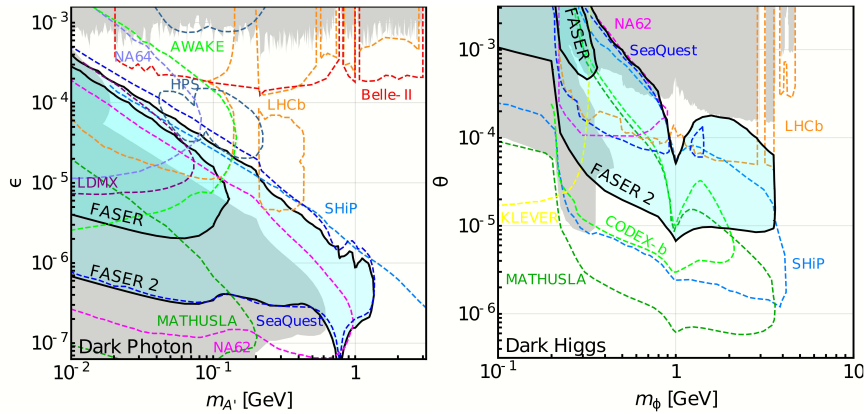}%
\end{center}
\caption{Exclusion plots for dark photon and dark Higgs searches: the two black areas correspond to the sensitivities of the FASER experiment in Run III and of the proposed FASER 2 upgrade in HL-LHC.
}
\label{fig:faser}
\end{figure}

The main scientific goal of the experiment is the search of new weakly-interacting light particles, which are produced mainly in the forward regions due to kinematic constraints. For example, we can look for a candidate event as the following: a very rare decay of a $\pi^{0}$ produced by collisions, like $\mathrm{\pi^{0} \rightarrow A' + X}$, where $\mathrm{A}$ is a dark photon; the dark photon reaches the detector and decay in a $e^{+}-e^{-}$ couple inside the decay volume; the event can be later identified by no signal in veto scintillators, two clear tracks in the spectrometer and an electromagnetic-like energy deposit in the calorimeter. To have an idea, with an aperture of 20 cm, corresponding to only $2\times10^{-8}$ of the solid angle, about 2\% of all $\pi^{0}$s produced in p-p collisions are inside the detector acceptance. In Run III, with an expected integrated luminosity of 150~$\mathrm{fb^{-1}}$, $10^{15}$ $\pi^{0}$s are directed towards the FASER detector, and the experiment has already a very good sensitivity to dark photon candidates. Fig.\ref{fig:faser} shows the experiment sensitivities to dark photons and dark Higgs considering all possible production and decay channels. As we can see, the experiment would strongly benefit from a possible upgrade after Run III, called FASER 2, increasing the decay volume to 5~m and the aperture to 2~m. In this way, with an expected integrated luminosity of 3~$\mathrm{ab^{-1}}$ in HL-LHC, the experiment would have excellent sensitivities to look for a wide range of BSM particles, like dark photons, dark Higgs, heavy neutral leptons and axion-like particles.

Another fundamental goal of the experiment is relative to neutrino physics. This is possible thanks to the dedicated FASER$\nu$ detector, installed in front of the FASER detector \cite{ref:faserni}, made of alternated layers of a couple of 50~$\mathrm{\mu m}$ emulsion films and 1~mm thick tungsten target, for a total mass of about 1.1~ton. In Run III, the experiment will measure neutrino production cross sections at TeV energies, separately for the three different lepton families. This will be not only the first ever measurement of neutrino production at a collider, but it will also involve an energy range where the production cross sections are not currently constrained. A first result obtained from a pilot run in 2018 with a reduced version of the detector was recently published, leading to six neutrino candidates produced in p-p collisions at $\sqrt{s} = $ 13 TeV for an integrated luminosity of 12.2~$\mathrm{fb^{-1}}$ \cite{ref:neutrino}.

\bibliographystyle{plain}
\bibliography{proc_berti_lhcp2021.bib}

%
%
%

\end{document}